\documentclass[12pt]{iopart}
\usepackage{graphicx}
\usepackage{iopams}
\usepackage{dcolumn}
\usepackage{bm}
\usepackage{amsthm}
\usepackage{amssymb}
\usepackage[flushleft]{caption}

\begin{document}

\title{Quantum Fisher Information of Entangled Coherent States in a Lossy Mach-Zehnder Interferometer}

\author{Xiaoxing Jing, Jing Liu, Wei Zhong, and Xiaoguang Wang}

\address{Zhejiang Institute of Modern Physics, Department of Physics, Zhejiang
University, Hangzhou 310027, China}
\ead{xgwang@zimp.zju.edu.cn}

\begin{abstract}
We give an analytical result for the quantum Fisher information of
entangled coherent States in a lossy Mach-Zehnder Interferometer
recently proposed by J. Joo $\emph{et al. }$[Phys. Rev. Lett. 107,
083601(2011)]. For small loss of photons, we find that the entangled
coherent state can surpass the Heisenberg limit. Furthermore, The
formalism developed here is applicable to the study of phase
sensitivity of multipartite entangled coherent states.
\end{abstract}

\submitto{\JPB}
\maketitle

\section{INTRODUCTION}
Precision measurements are important across all fields of science
and technology. By employing quantum features like entanglement and
squeezing, quantum metrology promises enhancing precision and has
drawn a lot of attention in the last decade ~\cite{giovannetti_quantum-enhanced_2004,giovannetti_quantum_2006,giovannetti_advances_2011,berry_how_2009,joo_quantum_2011,
joo_quantum_2012,bollinger_optimal_1996,luis_nonlinear_2004,boixo_quantum-limited_2008,
escher_quantum_2012,gendra_quantum_2013,gendra_quantum_2013,kasevich_measurement_1992,peters_measurement_1999,
wineland_squeezed_1994,wineland_spin_1992,wineland_squeezed_1994,santarelli_quantum_1999,chuang_quantum_2000,jozsa_quantum_2000,rivas_precision_2010}.
Quantum metrology deals with the ultimate precision limits in estimation
procedures, taking into account the constraints imposed by quantum mechanics, and
allows one to gain advantages over purely classical approaches~\cite{giovannetti_quantum-enhanced_2004,giovannetti_quantum_2006,giovannetti_advances_2011,berry_how_2009}. As a key component of the quantum metrology
theory, quantum parameter estimation has many applications in experiments, such as the detection
of gravitational radiation~\cite{kasevich_measurement_1992,peters_measurement_1999}, quantum frequency standards~\cite{wineland_spin_1992,wineland_squeezed_1994,santarelli_quantum_1999}, clock synchronization~\cite{chuang_quantum_2000,jozsa_quantum_2000}, to name a few.

Quantum Fisher information (QFI) is another significant concept in quantum
metrology and has been studied widely~\cite{braunstein_statistical_1994,Helstrom_quantum_detection_1976,Holevo_probabilistic_1982,dorner_optimal_2009,toth_multipartite_2012,pezze_entanglement_2009,
ma_fisher_2009,rivas_precision_2010,ma_quantum_2011,hubner_explicit_1992,zhong_fisher_2013}.
As an extension of the classical Fisher information in statistics
and information theory, QFI plays a paramount
role in quantum estimation theory. In quantum metrology theory, these
two concepts are linked by the quantum Cram\'er-Rao inequality~\cite{Helstrom_quantum_detection_1976,Holevo_probabilistic_1982},
\begin{equation}
{\rm var}(\hat{\varphi})\geq\frac{1}{\nu F},\label{eq:cr}
\end{equation}
where ${\rm var}(\hat{\varphi})$ is the variance of an
unbiased estimator $\hat{\varphi}$ of a parameter $\varphi$, $\nu$ represents
the number of repeated experiments and $F$ is the QFI of the parameter. The inverse of the QFI provides the
lower bound of the error of the estimation.

In this paper, we consider a fundamental parameter estimation task
in which the parameter $\varphi$ is generated by some unitary dynamics
$U=\exp(-i\varphi H)$. This kind of parameter estimation task is common
in many experimental setups such as Mach-Zehnder interferometers and Ramsey interferometers.
Based on a recent expression of QFI \cite{liujing_matrix_qfi_2013},
we show that the QFI of $\varphi$ for a unitary parameterized dynamics is the mean variance of $H$ over
the eigenstates minus the transition terms of $H$. Next we take
a two dimension case as our interest. The eigenvalues and eigenstates
of a general $2\times2$ density matrix have been given in terms of
its determinant, difference between diagonal elements and phase of
off-diagonal elements. For integrity we also give the eigenvalues and eigenstates
for a density operator on a nonorthogonal basis of two dimensions.

While exact results and analytical solutions are known for noiseless
situations, the determination of the ultimate precision limit in the
presence of noise is still a challenging problem in quantum mechanics.
Recently, J. Joo $\emph{et al.}$studied the entangled coherent states
in a Mach-Zehnder interferometer under perfect and lossy conditions \cite{joo_quantum_2011}.
They found the entangled coherent states (ECS) can reach better precision
in comparison to N00N, ``bat'', and ``optimal'' states in both
conditions. In lossy conditions, they modeled the particle loss by fictitious beam splitters and
adopted a numeric strategy to calculate the QFI of the ECS. Utilizing our formula we give an analytic
expression of the QFI. We find that even in a lossy condition, the ECS can still surpass the Heisenberg limit.

This paper is organized as follows. In Sec. II, we give a brief review
of the QFI and obtain an explicit formula of the QFI for a family of density matrices parameterized
through a unitary dynamics. In Sec. III, we give the eigenvalues and
eigenstates of a 2-dimensional density matrix in terms of its determinant,
difference between diagonal elements and phase of off-diagonal elements. We also generalize
the eigen problem in a nonorthogonal basis. Afterward,
in Sec. IV, we apply our result to the ECS in a lossy Mach-Zehnder interferometer and get an analytical expression
of the QFI. Finally, the conclusion is given in Sec. V.

\section{QFI AND PARAMETER ESTIMATION FOR UNITARY DYNAMICS }
\subsection{Brief Review of Quantum Fisher Information}

In this section, we briefly review the calculation of the QFI.
For a parameterized quantum states $\rho_{\varphi}$,
a widely used version of QFI $F_{\varphi}$ is defined as \cite{Helstrom_quantum_detection_1976,Holevo_probabilistic_1982}
\begin{equation}
F_{\varphi}:={\rm tr}(\rho_{\varphi}L^{2}),
\end{equation}
where the symmetric logarithmic derivative (SLD) operator  $L$ is determined
by
\begin{equation}
\partial_{\varphi}\rho_{\varphi}=\frac{1}{2}[L\rho_{\varphi}+\rho_{\varphi} L].\label{eq:sld}
\end{equation}

Consider a density operator $\rho_{\varphi}$ on a $N$-dimensional
system ($N$ can be infinite). The corresponding spectrum decomposition is given by
\begin{equation}
\rho_{\varphi}=\sum_{i=1}^{M}p_{i}|\psi_{i}\rangle\langle\psi_{i}|,
\end{equation}
where $p_{i}$ is the eigenvalue and $|\psi_{i}\rangle$ is the eigenstate,
and $M\leq N,$ implying that there are $N-M$ zero eigenvalues.
With the decomposition of the density matrix %$\rho_{\varphi}=\sum_{i}p_{i}|\psi_{i}\rangle\langle\psi_{i}|$,
one can directly obtain the element of the SLD operator from Eq.~(\ref{eq:sld}) as
\begin{equation}
\langle\psi_{k}|L|\psi_{l}\rangle=\frac{2\langle\psi_{k}|\partial_{\varphi}\rho_{\varphi}|\psi_{l}\rangle}{p_{l}+p_{k}}.
\end{equation}
Notice that the matrix element of $L$ is not defined when $p_{l}+p_{k}=0$.

It turns out that the QFI is completely determined in
the support of $\rho_{\varphi}$, that is, the space spanned by those eigenvectors
corresponding to the nonvanishing eigenvalues. It can be expressed as \cite{liujing_matrix_qfi_2013}
\begin{eqnarray}
F_{\varphi} & = & \sum_{i=1}^{M}\frac{1}{p_{i}}(\partial_{\varphi}p_{i})^{2}+\sum_{i=1}^{M}4p_{i}\langle\partial_{\varphi}\psi_{i}|\partial_{\varphi}\psi_{i}\rangle\nonumber \\
 &  & -\sum_{i=1}^{M}\sum_{j=1}^{M}\frac{8p_{i}p_{j}}{(p_{i}+p_{j})}|\langle\psi_{i}|\partial_{\varphi}\psi_{j}\rangle|^{2}.\label{eq:QFI1}
\end{eqnarray}

For the special case of a pure state ($M=1$), Eq.~(\ref{eq:QFI1}) reduces
to
\begin{equation}
F(\psi_{1})=4[\langle\partial_{\varphi}\psi_{1}|\partial_{\varphi}\psi_{1}\rangle-|\langle\psi_{1}|\partial_{\varphi}\psi_{1}\rangle|^{2}].
\end{equation}
Using this form of the QFI for pure states, we can
rewrite Eq.~(\ref{eq:QFI1}) as

\begin{eqnarray}
F_{\varphi} & = & \sum_{i=1}^{M}\frac{1}{p_{i}}(\partial_{\varphi}p_{i})^{2}+\sum_{i=1}^{M}p_{i}F(\psi_{i})\nonumber \\
 &  & -\sum_{i\neq j}^{M}\frac{8p_{i}p_{j}}{(p_{i}+p_{j})}|\langle\psi_{i}|\partial_{\varphi}\psi_{j}\rangle|^{2}.\label{eq:QFI2}
\end{eqnarray}
It is clear that the first term can be regarded as the classical contribution~\cite{Paris,liujing_matrix_qfi_2013,Holevo_probabilistic_1982}, and the second term as the
mean QFI over the eigenstates. The third term can be regarded as a sum of harmonic mean of transition terms.

There are several similar formulas in the literature where the summation in the last term runs over all the eigenstates,
as long as~$p_{i}+p_{j}\neq0$. Eq.~(\ref{eq:QFI2}) have some advantages over them both in analytical and numerical
calculations since $i,j$ are symmetric and one only need to find the non-varnishing eigenstates of $\rho_{\varphi}$.

\subsection{QFI for  unitary parameterized dynamics}

In quantum estimation theory, the most fundamental parameter estimation
task is to estimate a small parameter $\varphi$ generated by some
unitary dynamics
\begin{equation}
U=\exp(-i\varphi H).
\end{equation}
Here $H$ is a Hermitian operator and can be regarded as the generator of
parameter $\varphi$. This form of parameterization process is typical in
interferometers. For instance, in a Ramsey interferometer $H$ can be a collective angular momentum operator
$J_{n}$~\cite{ma_quantum_2011}, which can be viewed as a generator of SU(2).
In Mach-Zehnder interferometers, denoting $a_{i}$ and $a_{i}^\dagger$ (i=1,2) as the annihilation and creation operators
for \emph{i}th mode, then $H$ can be (1) the photon number difference between two modes: $a_{1}^\dagger a_{1}-a_{2}^\dagger a_{2}$~\cite{jarzyna_quantum_2012},
(2) the number operator in one mode: $a_{2}^\dagger a_{2}$~\cite{joo_quantum_2011,jozsa_quantum_2000},
(3) the number operator to the \emph{k}$\rm{th}$ power: $(a_{2}^\dagger a_{2})^k$,
in a nonlinear interferometer~\cite{joo_quantum_2012}.

Suppose the initial state $\rho_{0}$ has already been
decomposed as
\begin{equation}
\rho_{0}=\sum_{i}^{M}p_{i}|\phi_{i}\rangle\langle\phi_{i}|.
\end{equation}
Here we assume $\rho_{0}$ is independent of $\varphi$.
After the unitary rotation, $\rho_{\varphi}$ can be decomposed as
\begin{eqnarray}
\rho_{\varphi}=\sum_{i}p_{i}|\psi_{i}\rangle\langle\psi_{i}|, \label{eq:decomposition}
\end{eqnarray}
with
\begin{equation}
|\psi_{i}\rangle=e^{-i\varphi H}|\phi_{i}\rangle.
\end{equation}

Substituting Eq.~(\ref{eq:decomposition}) into Eq.~(\ref{eq:QFI2}) leads to
the QFI given by
\begin{equation}
F_{\varphi}=4\left[\sum_{i=1}^{M}p_{i}(\Delta H_{i})^{2}-\sum_{i\neq j}^{M}\frac{2p_{i}p_{j}}{p_{i}+p_{j}}|H_{ij}|^{2}\right], \label{eq:QFI3}
\end{equation}
where
\begin{equation}
(\Delta H_{i})^{2}=\langle\phi_{i}|H^{2}|\phi_{i}\rangle-\langle\phi_{i}|H|\phi_{i}\rangle^{2}, \nonumber
\end{equation}
and
\begin{equation}
|H_{ij}|^{2}=|\langle\phi_{i}|H|\phi_{j}\rangle|^{2}, \nonumber
\end{equation}
are the variance and transition probability of $H$ in the eigenstates
of $\rho_{0}$. Since $p_{i}$ is independent of $\varphi$, the classical contribution vanishes. The first term in Eq.~(\ref{eq:QFI3}) is the mean variance of $H$ over the
eigenstates, while the second term is a sum of transition probability of $H$ with a harmonic
mean weight.

If~$\rho_{0}$~is a pure state, we can take $p_{i}=\delta_{i1,}$ then
\begin{equation}
F_{\varphi}=4(\Delta H_{1})^{2};
\end{equation}
if~$\rho_{0}$~only has two nonzero components, we take $p_{1}p_{2}\neq0$
and $p_{i}=0$ when $i>2$, then
\begin{equation}
F_{\varphi}=4p_{1}(\Delta H_{1})^{2}+4p_{2}(\Delta H_{2})^{2}-16p_{1}p_{2}|H_{12}|^{2}.\label{eq:qfi2d}
\end{equation}
In the following, we take $M=2$ as our main interest.

\section{EIGEN PROBLEM OF A Nonorthogonal $2\times2$ Density Matrix }

%\subsection{Eigenvalues and Eigenstates of A Nonorthogonal $2\times2$ Density Matrix }

According to Eq.~(\ref{eq:QFI2}) and Eq.~(\ref{eq:QFI3}), we only need to find the non-vanishing eigenstates of the density operator
rather than all its eigenstates. However, it is generally not feasible to get the analytical diagonalization of~$\rho_{\varphi}$. In that case,
one has to resort to numeric methods or decompose the density operator into a nonorthogonal basis and use the convexity of QFI.

In this paper, we develop a systematic routine to find the eigenvalues and eigenstates of a density operator of rank 2 and apply it to an
interesting scenario.
Let us consider a $2\times2$ density operator $\tilde{\rho}$
on a $\emph{nonorthogonal}$ basis
\begin{equation}
\tilde{\rho}=a|\Psi_{1}\rangle\langle\Psi_{1}|+b|\Psi_{1}\rangle\langle\Psi_{2}|+b^{*}|\Psi_{2}\rangle\langle\Psi_{1}|+d|\Psi_{2}\rangle\langle\Psi_{2}|,\label{eq:NonorthRho}
\end{equation}
where $|\Psi_{1}\rangle,|\Psi_{2}\rangle$ are normalized states and
$a,d$ are real numbers due to the hermiticity of density operator.
The special case when $|\Psi_{1}\rangle$ and $|\Psi_{2}\rangle$ are orthogonal
is discussed in Appendix A. In order to get the eigenvalues and eigenvectors of $\tilde{\rho}$,
we first recast it into an orthogonal basis (one can also solve the
eigen problem in the original nonorthogonal basis, see Appendix B.)

Denoting $p=\langle\Psi_{1}|\Psi_{2}\rangle$, we introduce a new
set of basis by the Gram-Schmidt procedure \cite{Nielsen_quantum_computation_2000}
\begin{eqnarray}
|\Phi_{1}\rangle & = & |\Psi_{1}\rangle,\nonumber \\
|\Phi_{2}\rangle & = & \frac{1}{\sqrt{1-|p|^{2}}}(|\Psi_{2}\rangle-p|\Psi_{1}\rangle),\nonumber \label{eq:NewBasis}
\end{eqnarray}
which are orthonormal.
Through the inverse transformation: $|\Psi_{1}\rangle=|\Phi_{1}\rangle$,
$|\Psi_{2}\rangle=\sqrt{1-|p|^{2}}|\Phi_{2}\rangle+p|\Phi_{1}\rangle$,
the density matrix in the new basis reads
\begin{equation}
\tilde{\rho}=\left(\begin{array}{cc}
a+bp^{*}+b^{*}p+d|p|^{2} & (b+dp)\sqrt{1-|p|^{2}}\\
(b^{*}+dp^{*})\sqrt{1-|p|^{2}} & d(1-|p|^{2})
\end{array}\right).\label{eq:RhoNewBasis}
\end{equation}
The determinant of this density matrix, expectation value of $\sigma_{3}$
and off-diagonal phase read
\begin{eqnarray}
{\rm det}(\tilde{\rho}) & = & (1-|p|^{2})(ad-|b|^{2}),\label{eq:detrhotilde} \nonumber\\
\langle\sigma_{3}\rangle_{\tilde{\rho}} & = & 1-2d(1-|p|^{2}),\label{eq:sigma3} \nonumber\\
e^{i\tilde{\tau}} & = & \frac{b+dp}{|b+dp|}.\label{eq:taotilde}
\end{eqnarray}

According to appendix A, the eigenvalues and eigenstates of $\tilde{\rho}$
can be expressed in terms of $\rm{det}(\tilde{\rho}),\langle\sigma_{3}\rangle_{\tilde{\rho}}$ and $\tilde{\tau}$.
For clarity, we denote the eigenvalues and eigenstates
as $\tilde{\lambda}_{\pm}$ and $|\tilde{\lambda}_{\pm}\rangle$
correspondingly. The values of $\tilde{\lambda}_{\pm}$ are
\begin{equation}
\tilde{\lambda}_{\pm}=\frac{1\pm\sqrt{1-4{\rm det}(\tilde{\rho})}}{2}, \label{eq:eigenvalue}
\end{equation}
and the eigenstates read
\begin{eqnarray}
|\tilde{\lambda}_{+}\rangle & = & \tilde{v}_{+}e^{i\tilde{\tau}}|\Phi_{1}\rangle+\tilde{v}_{-}|\Phi_{2}\rangle, \nonumber \\
|\tilde{\lambda}_{-}\rangle & = & -\tilde{v}_{-}e^{i\tilde{\tau}}|\Phi_{1}\rangle+\tilde{v}_{+}|\Phi_{2}\rangle,\label{eq:EigenstateinPhi}
\end{eqnarray}
where
\begin{equation}
\tilde{v}_{\pm}  =  \left(\frac{\sqrt{1-4{\rm det}(\tilde{\rho})}\pm\langle\sigma_{3}\rangle_{\tilde{\rho}}}{2\sqrt{1-4{\rm det}(\tilde{\rho})}}\right)^{\frac{1}{2}}.   \label{eq:vpm}
\end{equation}
Hence the density matrix can be decomposed as
\begin{equation}
\tilde{\rho}=\sum_{i=\pm}\tilde{\lambda}_{i}|\tilde{\lambda}_{i}\rangle\langle\tilde{\lambda}_{i}|.
\end{equation}
Alternatively, one can transform the eigenstates back to the nonorthogonal
basis,
\begin{eqnarray}
|\tilde{\lambda}_{+}\rangle&=&(\tilde{v}_{+}e^{i\tilde{\tau}}-\frac{p\tilde{v}_{-}}{\sqrt{1-|p|^{2}}})|\Psi_{1}\rangle
+\frac{\tilde{v}_{-}}{\sqrt{1-|p|^{2}}}|\Psi_{2}\rangle,\nonumber \\
\label{eq:eigenstatesinPsi1}  \nonumber\\
|\tilde{\lambda}_{-}\rangle&=&(-\tilde{v}_{-}e^{i\tilde{\tau}}-\frac{p\tilde{v}_{+}}{\sqrt{1-|p|^{2}}})|\Psi_{1}\rangle
+\frac{\tilde{v}_{+}}{\sqrt{1-|p|^{2}}}|\Psi_{2}\rangle. \nonumber\\
 \label{eq:eigenstatesinPsi2}
\end{eqnarray}

\section{QFI OF ECS IN A LOSSY MACH-ZEHNDER INTERFEROMETER}

\subsection{Reformulation of the Density Matrix of ECS in A Lossy Mach-Zehnder
Interferometer}

In a recent paper~\cite{joo_quantum_2011}, the author analyzed
the QFI of an entangled coherent state(ECS) in the Mach-Zehnder interferometer. The main idea of their proposition
is as follows. A coherent state $|\alpha/\sqrt{2}\rangle$
and a coherent state superposition(CSS)
\begin{equation}
|\rm{CSS}\rangle=\mathcal{N}_{\alpha}(|\frac{\alpha}{\sqrt{2}}\rangle+|\frac{-\alpha}{\sqrt{2}}\rangle),
\end{equation}
are fed into the first 50:50 beam splitter of the Mach-Zehnder interferometer and become the ECS,
\begin{equation}
|\rm{ECS}\rangle_{1,2}=\mathcal{N}_{\alpha}[|\alpha\rangle_{1}|0\rangle_{2}+|0\rangle_{1}|\alpha\rangle_{2}],
\end{equation}
where
\begin{equation}
\mathcal{N}_{\alpha}=1/\sqrt{2(1+e^{-|\alpha|^{2}}})
\end{equation}
is the normalized coefficient.
Then a parameter is imprinted in one of the mode by a unitary phase shift $U(\varphi)$.
They modeled particle loss in the realistic scenario by two fictitious
beam splitters $B_{1,3}^{T},$ $B_{2,4}^{T}$ with the same transmission coefficient T.
When $T=1$, the interferometer has no photon loss.
Here the subscript $3,4$ represent the environment modes.
After tracing out the environment modes, they got the density matrix of the original
mode $\rho_{12}$.

To calculate the QFI of $\rho_{12}$,
they adopted numerical methods and truncated the coherent state at $n=15$.
Using the approach developed in Sec.~(II) and Sec.~(III), we can give the analytical expression of the
QFI. In the following, we reformulate the density operator in a form as Eq.~(\ref{eq:NonorthRho}).

First, we denote the density operator before phase shift and particle loss as
\begin{equation}
\rho_{\rm{in}}=|\rm{ECS}\rangle_{1,2}|0\rangle_{3}|0\rangle_{4}\langle0|_{4}\langle0|_{3}\langle\rm{ECS}|_{1,2}.
\end{equation}
In the interferometer, $\rho_{\rm{in}}$ suffers both particle loss and phase shift before exiting the second
50:50 beam splitter. The phase accumulation $U(\varphi)=e^{-i\varphi a_{2}^{\dagger}a_{2}}$
and the particle loss process, indicated by the fictitious beam splitters $B_{1,3}^{T}$, $B_{2,4}^{T}$, are
interchangeable~\cite{dorner_optimal_2009,demkowicz-dobrzanski_quantum_2009}. Here $B_{1,3}^{T}$ and $B_{2,4}^{T}$ satisfy the relation~\cite{wang_bipartite_2002}
\begin{equation*}
B^{T}_{1,2}|\alpha_{1}\rangle_{1}|\alpha_{2}\rangle_{2}
=|\alpha_{1}\sqrt{T}+\alpha_{2}\sqrt{R}\rangle_{1}|\alpha_{1}\sqrt{R}
-\alpha_{2}\sqrt{T}\rangle_{2}.
\end{equation*}
Thus the
final reduced density operator can be written as
\begin{eqnarray}
\rho_{1,2} & = & {\rm Tr_{3,4}}(B_{1,3}^{T}B_{2,4}^{T}U\rho_{\rm{in}}U^{\dagger}B_{2,4}
^{T\dagger}B_{1,3}^{T\dagger}) \label{temp1} \\
& = & {\rm {\rm Tr_{3,4}}}(UB_{1,3}^{T}B_{2,4}^{T}\rho_{\rm{in}}B_{2,4}^{T\dagger}B_{1,3}^{T\dagger}U^{\dagger}) \label{temp2}.
\end{eqnarray}
The authors in Ref.~\cite{joo_quantum_2011} use the expression (\ref{temp1}).
To apply our result in Sec.~(II) and Sec.~(III), we take the expression (\ref{temp2}).

Second, the phase accumulation operator can be brought forward further, i.e.,
\begin{eqnarray}
\rho_{1,2} & = & U{\rm Tr_{3,4}}(B_{1,3}^{T}B_{2,4}^{T}\rho_{\rm{in}}B_{2,4}^{T\dagger}B_{1,3}^{T\dagger})U^{\dagger}\nonumber \\
 & = & U\tilde{\rho}_{1,2}U^{\dagger},
\end{eqnarray}
where
\begin{equation}
\tilde{\rho}_{1,2} = {\rm Tr_{3,4}}(B_{1,3}^{T}B_{2,4}^{T}\rho_{\rm{in}}B_{2,4}^{T\dagger}B_{1,3}^{T\dagger}).
\end{equation}
That is, in such a lossy situation, the phase shift is still a unitary process for $\tilde{\rho}_{1,2}$. Therefore we can calculate the QFI of $\rho_{1,2}$ by
finding the decomposition of $\tilde{\rho}_{1,2}$.
With the denotation of
\begin{eqnarray*}
\alpha'&=& \alpha\sqrt{T}   \\
\beta' &=& \alpha\sqrt{1-T}=\alpha\sqrt{R}
\end{eqnarray*}
$\tilde{\rho}_{1,2}$ can be specifically calculated as
\begin{eqnarray}
\tilde{\rho}_{1,2} & = & \mathcal{N}_{\alpha}^{2}[|\alpha',0\rangle\langle\alpha',0|+e^{-|\beta'|^{2}}|\alpha',0\rangle\langle0,\alpha'|\nonumber \\
& & +e^{-|\beta'|^{2}}|0,\alpha'\rangle\langle\alpha',0|+|0,\alpha'\rangle\langle0,\alpha'|].\label{eq:Rho12}
\end{eqnarray}
We can see $\tilde{\rho}_{1,2}$ has the same form of Eq.~(\ref{eq:NonorthRho}). In the next subsection we show the decomposition of $\tilde{\rho}_{1,2}$ and calculate the QFI.

\subsection{Calculation of the ECS's QFI}

In order to find the decomposition of $\tilde{\rho}_{1,2}$, we set $|\Psi_{1}\rangle=|\alpha',0\rangle,$
$|\Psi_{2}\rangle=|0,\alpha'\rangle$ correspondingly. Comparing Eq.~(\ref{eq:Rho12}) with
Eq.~(\ref{eq:NonorthRho}),
we can find the determinant of this density matrix, expectation value of $\sigma_{3}$ and off-diagonal phase as
\begin{eqnarray}
{\rm det}(\tilde{\rho}_{1,2}) &=& \mathcal{N}_{\alpha}^{4}(1-e^{-2|\alpha^{'}|^{2}})(1-e^{-2|\beta^{'}|^{2}}),\label{eq:detrho12s} \nonumber\\
\langle\sigma_{3}\rangle_{\tilde{\rho}_{1,2}} &=& 1-2\mathcal{N}_{\alpha}^{2}+2\mathcal{N}_{\alpha}^{2}e^{-2|\alpha^{'}|^{2}},\label{eq:sigma3s} \nonumber\\
e^{i\tilde{\tau}} &=& 1. \label{eq:tildetaus}
\end{eqnarray}
According to the preceding section, we can find the eigenvalues as
\begin{equation}
\tilde{\lambda}_{\pm}=\frac{1}{2}\pm\frac{\sqrt{2e^{-|\alpha|^{2}}+e^{-2|\alpha'|^{2}}+e^{-2|\beta'|^{2}}}}{2+2e^{-|\alpha|^{2}}},
\end{equation}
and
\begin{equation}
\tilde{v}_{\pm}=\frac{1}{2}\pm\frac{e^{-|\alpha|^{2}}+e^{-2|\alpha'|^{2}}}{2\sqrt{2e^{-|\alpha|^{2}}+e^{-2|\alpha'|^{2}}+e^{-2|\beta'|^{2}}}}.
\end{equation}

Next we analyze the parametrization procedure. The unitary operator on
$\tilde{\rho}_{1,2}$ reads
\begin{equation}
U(\varphi)=\exp(-i\varphi a_{2}^{\dagger}a_{2}),
\end{equation}
i.e., the generator of $\varphi$ is $H=a_{2}^{\dagger}a_{2}$. According to Eq.~(\ref{eq:qfi2d}),
we only need to calculate the variance of $H$ in $|\tilde{\lambda}_{\pm}\rangle$
and the transition probability of $H$
between $|\tilde{\lambda}_{\pm}\rangle$. Since $H|\Psi_{1}\rangle=0$,
we choose Eq.~(\ref{eq:eigenstatesinPsi2}) for convenience.

The variance in $|\tilde{\lambda}_{+}\rangle$ is
\begin{eqnarray}
\Delta H_{1}^{2} & = & \langle\tilde{\lambda}_{+}|(a_{2}^{\dagger}a_{2})^{2}|\tilde{\lambda}_{+}\rangle
-(\langle\tilde{\lambda}_{+}|a_{2}^{\dagger}a_{2}|\tilde{\lambda}_{+}\rangle)^{2}\nonumber \\
& = & \frac{\tilde{v}_{-}^{2}}{1-p^{2}}(|\alpha'^{2}|^{2}+|\alpha'|^{2}-\frac{\tilde{v}_{-}^{2}}{1-p^{2}}|\alpha'|^{4}).\label{eq:variance1}
\end{eqnarray}
Similarly, the variance in $|\tilde{\lambda}_{-}\rangle$ is
\begin{equation}
\Delta H_{2}^{2}=\frac{\tilde{v}_{+}^{2}}{1-p^{2}}(|\alpha'^{2}|^{2}+|\alpha'|^{2}-\frac{\tilde{v}_{+}^{2}}{1-p^{2}}|\alpha'|^{4}),\label{eq:variance2}
\end{equation}
and the transition term is
\begin{equation}
|H_{12}|^{2}  =  (\frac{\tilde{v}_{+}\tilde{v}_{-}}{1-p^{2}}|\alpha'|^{2})^{2}.\label{eq:transition}
\end{equation}

Utilizing above expressions and based on Eq.~(\ref{eq:qfi2d}),
we can obtain the QFI of $\rho_{1,2}$ as
\begin{equation}
F=4\mathcal{N}_{\alpha}^{2}|\alpha|^{2}T\left[1+\mathcal{G}(T,\alpha)\right],  \label{eq:fa}
\end{equation}
where
\begin{equation*}
\mathcal{G}(T,\alpha)=\frac{
(\mathcal{N}_{\alpha}^{2}-1)e^{-2|\alpha|^{2}T}+\mathcal{N}_{\alpha}^{2}e^{-2|
\alpha|^{2}R}+2\mathcal{N}_{\alpha}^{2}e^{-|\alpha|^{2}}
}{1-e^{-2|\alpha|^{2}T}}|\alpha|^{2}T.
\end{equation*}
Notice that $\mathcal{N}_{\alpha}$ satisfies the relation
\begin{equation*}
2\mathcal{N}_{\alpha}^{2}e^{-|\alpha|^2}=1-2\mathcal{N}_{\alpha}^{2},
\end{equation*}
then $\mathcal{G}(T,\alpha)$ can be rewritten as
\begin{equation*}
\mathcal{G}(T,\alpha) = |\alpha|^{2}T \left[1-\mathcal{N}_{\alpha}^{2}-
\frac{\mathcal{N}_{\alpha}^{2}(1-e^{-2|\alpha|^{2}R})}
{1-e^{-2|\alpha|^{2}T}}\right].
\end{equation*}
Introduce the total average photon number $\bar{n}=\langle a_{1}^{\dagger}a_{1}+a_{2}^{\dagger}a_{2}\rangle$,
and it is easy to find that in this case
\begin{equation*}
\bar{n}=2\mathcal{N}_{\alpha}^{2}|\alpha|^2,
\end{equation*}
then $\mathcal{G}(T,\alpha)$ can be further written into
\begin{equation}
G(T,\alpha)= T\left[|\alpha|^{2}-\frac{\bar{n}}{2}-
\frac{\bar{n}}{2}\frac{1-e^{-2|\alpha|^{2}R}}
{1-e^{-2|\alpha|^{2}T}}\right],
\end{equation}
and the QFI (\ref{eq:fa}) can be finally simplified as
\begin{equation}
F=\bar{n}T\left[2+\left(2|\alpha|^{2}-\bar{n}-\bar{n}\frac{1-e^{-2|\alpha|^{2}R}}
{1-e^{-2|\alpha|^{2}T}}\right)T\right]. \label{eq:fa2}
\end{equation}
The QFI is only determined by the total average photon number $\bar{n}$
and the transmission coefficient $T$.

When ~$T=R=1/2$~, the QFI reduces into
\begin{eqnarray}
F=\bar{n}+\frac{\bar{n}}{2}(|\alpha|^{2}-\bar{n})\geq\bar{n}.
\end{eqnarray}
The last inequality is due to the fact that~$|\alpha|^{2}\geq\bar{n}$~with the equal sign holds in the limit of~$|\alpha|^{2}\rightarrow\infty$~.
Since $F$ decreases monotonically with the transmission coefficient, the ECS can surpass the shot noise limit as long as $T>\frac{1}{2}$;
when~$T=1-R=1$~, i.e., there is no particle loss in the interferometer, the QFI can be simplified as
\begin{equation}
F=\bar{n}\left(2+2|\alpha|^{2}-\bar{n}\right),
\end{equation}
and due to~$|\alpha|^{2}\geq\bar{n}$~, we have
\begin{equation}
F\geq\bar{n}^{2}+2\bar{n}.  \label{eq:QFIpure}
\end{equation}

There is a debate over the ultimate scaling of the phase sensitivity for states with a fluctuating number of particles~\cite{hyllus_entanglement_2010}.
There are two candidates in the literature: the so-called Hofmann limit~$\delta\varphi\sim1/\sqrt{\overline{n^{2}}}$, and the Heisenberg limit $\delta\varphi\sim1/\overline{n}$.
Here we will show that the ECS can surpass the Heisenberg limit and Hofmann limit, even in the presence of particle loss.

From inequality~(\ref{eq:QFIpure}), one can find that the QFI without particle loss is greater than $\overline{n}^{2}$, next we will show it is also greater than $\overline{n^{2}}$.
The average of $n^{2}=(a_{1}^{\dagger}a_{1}+a_{2}^{\dagger}a_{2})^{2}$ does not change after the first beam splitter. Then it is easy to find
\begin{eqnarray*}
\overline{n^{2}} &=& \langle\mathrm{ECS}|_{1,2}(a_{1}^{\dagger}a_{1}+a_{2}^{\dagger}a_{2})^{2}
|\mathrm{ECS}\rangle_{1,2}  \\
&=& 2\mathcal{N}_{\alpha}^{2}\left[|\alpha|^{2}+|\alpha|^{4}\right] \\
&=& \left(1+|\alpha|^{2}\right)\overline{n},
\end{eqnarray*}
and compare with the QFI, we have
\begin{eqnarray}
F=2\overline{n^{2}}-\overline{n}^{2}=\overline{n^{2}}+\Delta(n), \label{eq:qfipure2}
\end{eqnarray}
where $\Delta(n)$ is the variance of the photon number. It is clear that $F$ is larger than both of $\overline{n^{2}}$ and $\overline{n}^{2}$.

\begin{figure}[tbp]
\centering\includegraphics[width=8.5cm,clip]{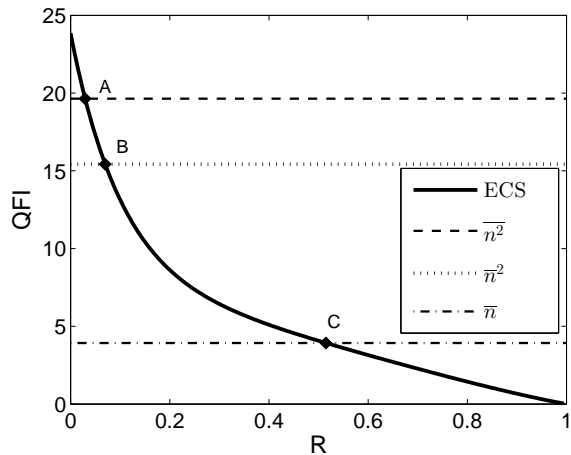}
\caption{The QFI of ECS with particle loss. Here $R=1-T$, $|\alpha|=2$. When the particle loss
is small, the QFI is larger than both $\overline{n^{2}}$ and $\overline{n}^{2}$.}
\label{fig2}
\end{figure}

Figure.~\ref{fig2} shows the variation of QFI with the increase of $R$. Points A, B and C represent the intersection with the Hofmann limit, Heisenberg limit and shot noise limit repectively. The corresponding reflection coefficients read $R_{\mathrm{A}}=0.03$, $R_{\mathrm{B}}=0.07$ and $R_{\mathrm{C}}=0.52$. From this figure, one can find that when $R<R_{\mathrm{A}}$, the ECS can always surpass the Hofmann limit, and for $R<R_{\mathrm{B}}$, the precision is still better than the Heisenberg limit. This indicates that the precision is robust and overcomes the Heisenberg limit with a small loss of photons within $R_{\mathrm{B}}$.
If the precision is only required in the range of shot noise limit, then this interferometer can tolerate a loss of half photons.

The ECS is very useful and robust for quantum metrology \cite{munro_weak-force_2002,ralph_coherent_2002}. Our formula gives an easy
approach to the determination of the QFI of ECS and one doesn't have to resort to numeric
methods.

\section{Conclusion}

We have derived an explicit formula for the QFI for a large class of
states in which the parameter is introduced by a unitary dynamics
$U=\exp(-iH\varphi)$. We pointed out that the QFI in this scenario
is the mean variance of $H$ over the eigenstates minus weighted
cross terms. Finally, we analyzed the QFI of a density matrix with
$M=2$ and apply our result into an entangled coherent state in a
Mach-Zehnder interferometer, which was proposed in a recent paper
\cite{joo_quantum_2011}.

We have found the analytical expression of the QFI for the ECS when
there is particle loss. We find that even in the lossy condition,
the ECS can still surpass the Heisenberg limit. The formalism
developed here can be applicable to the study of more complicated
states, such as the reduced two-mode mixed state when the total
multi-mode system is in a multipartite entangled coherent states.

\ack
The authors thank Xiao-Ming Lu and  Qing-Shou Tan for useful discussion.
This work was supported by the NFRPC with Grant No.2012CB921602 and NSFC with Grant No.11025527 and No.10935010.

\emph{Note added}: After the submission of our manuscript, we notice that the authors in Ref.~\cite{Jin} do a relevant work and have a similar conclusion.

\appendix
\section{Eigenvalues and Eigenstates of A $2\times2$ Density Matrix }

A general $2\times2$ density matrix $\rho$ is given in the form

\begin{equation}
\rho=\left(\begin{array}{cc}
\eta & \xi e^{i\tau}\\
\xi e^{-i\tau} & 1-\eta
\end{array}\right).\label{eq:GeneralRho}
\end{equation}
For this matrix to represent a physical state, one condition must
be met: the determinant of $\rho$ must be positive, i.e., ${\rm det}(\rho)=\eta(1-\eta)-\xi^{2}\geq0$
(this inequality implies $\eta\geq0$ , thus fullfil the positivity
requirement of density matrix). Here $\xi>0$, $\tau\in[0,2\pi)$
are real numbers due to the Hermiticity of density matrix.

The eigenvalues of $\rho$ can be easily calculated as
\begin{eqnarray}
\lambda_{\pm} & = & \frac{1\pm\sqrt{1-4{\rm det}(\rho)}}{2},\label{eq:EigValofGeneralRho}
% & = & \frac{1\pm\sqrt{\langle\sigma_{3}\rangle^{2}+4\xi^{2}}}{2},
\end{eqnarray}
and the corresponding normalized eigenvectors read
\begin{eqnarray}
|\lambda_{+}\rangle&=&\left(v_{+}e^{i\tau},v_{-}\right)^{\rm{T}}, \nonumber  \\
|\lambda_{-}\rangle&=&\left(-v_{-}e^{i\tau},v_{+}\right)^{\rm{T}},
\end{eqnarray}
with
\begin{eqnarray}
v_{\pm}  =  \left(\frac{\sqrt{1-4{\rm det}(\rho)}\pm\langle\sigma_{3}\rangle}{2\sqrt{1-4{\rm det}(\rho)}}\right)^{\frac{1}{2}},
%v_{2} & = & \left(\frac{\sqrt{1-4{\rm det}(\rho)}-\langle\sigma_{3}\rangle}{2\sqrt{1-4{\rm det}(\rho)}}\right)^{\frac{1}{2}}.
\end{eqnarray}
Here $\sigma_{3}$ is a Pauli matrix
%=\left(\begin{array}{cc}
%1 & 0\\
%0 & -1
%\end{array}\right)$
and $\langle\sigma_{3}\rangle={\rm Tr}(\rho\sigma_{3})=2\eta-1$.

We can see that the eigenvalues and eigenvectors of $\rho$ are fully
determined by ${\rm det}(\rho),$ $\langle\sigma_{3}\rangle$ and
$\tau.$

\section{An equivalent way to solve the eigen problem of density operator in nonorthogonal basis }
In this appendix we provide an equivalent way to solve the eigen problem
of Eq.~(\ref{eq:NonorthRho}). Instead of recasting $\tilde{\rho}$ into an orthonormal
basis, we assume the eigenvector as
\begin{equation}
|\phi\rangle=c_{1}|\Psi_{1}\rangle+c_{2}|\Psi_{2}\rangle.
\end{equation}
Then the eigen equation reads
\begin{equation}
\tilde{\rho}|\phi\rangle=\lambda|\phi\rangle,
\end{equation}
specifically (in the basis of $|\Psi_{1,2}\rangle$),
\begin{equation}
\left(\begin{array}{cc}
a+bp^{*} & ap+b\\
b^{*}+cp^{*} & b^{*}p+c
\end{array}\right)\left(\begin{array}{c}
c_{1}\\
c_{2}
\end{array}\right)=\lambda\left(\begin{array}{c}
c_{1}\\
c_{2}
\end{array}\right),
\end{equation}
i.e., we need find the eigenvalues and eigenvectors of the left matrix.
One can easily find the trace and determinant are the same as those
of Eq.~(\ref{eq:RhoNewBasis}), thus the eigenvalues are equal according
to Eq.~(\ref{eq:EigValofGeneralRho}).

The eigenvectors can also be easily calculated as
\begin{eqnarray}
|\phi_{1}\rangle & = & P_{11}|\Psi_{1}\rangle+P_{21}|\Psi_{2}\rangle,\label{eq:EigenState1'} \nonumber\\
|\phi_{2}\rangle & = & P_{12}|\Psi_{1}\rangle+P_{22}|\Psi_{2}\rangle,\label{eq:EigenState2'}
\end{eqnarray}
with the normalized conditions
\begin{eqnarray}
|P_{11}|^{2}+|P_{21}|^{2}+2{\rm {Re}(pP_{11}^{*}P_{21})=1}, \nonumber  \\
|P_{12}|^{2}+|P_{22}|^{2}+2{\rm {Re}(pP_{12}^{*}P_{22})=1}, \label{eq:normalize}  \\ \nonumber
\end{eqnarray}
where ${\rm Re}$ stands for real component. After some straightforward calculation, we can find
\begin{eqnarray}
P_{11}&=&\tilde{v}_{+}e^{i\tilde{\tau}}-\frac{\tilde{v}_{-}p}{\sqrt{1-|p|^{2}}}, \nonumber\\
P_{21}&=&\frac{\tilde{v}_{-}}{\sqrt{1-|p|^{2}}},  \nonumber\\
P_{12}&=&-\tilde{v}_{-}e^{i\tilde{\tau}}-\frac{\tilde{v}_{+}p}{\sqrt{1-|p|^{2}}}, \nonumber\\
P_{22}&=&\frac{\tilde{v}_{+}}{\sqrt{1-|p|^{2}}},  \\ \nonumber
\end{eqnarray}
where $e^{i\tilde{\tau}}$ and $\tilde{v}_{\pm}$  are defined in Eq.~(\ref{eq:taotilde}) and Eq.~(\ref{eq:vpm}), i.e., the
eigenstates in Eq.~(\ref{eq:EigenState2'})
are actually the same with Eq.~(\ref{eq:eigenstatesinPsi2}).

This method is a routine way to solving eigen problem. However, taking account of the normalization condition Eq.~(\ref{eq:normalize}),
it is quite tedious in calculation. We hope the method in the main text can offer some convenience when dealing
with similar problems.

\end{document}